\documentclass[10pt,conference]{IEEEtran}
 \usepackage{amsmath,amssymb,graphicx}
\usepackage{epsfig,times}
\usepackage{graphics,color,pstricks,pst-grad}
\usepackage{citesort}  

\newtheorem{cor}{Corollary}[section]
\newtheorem{prop}{Proposition}[section]

\newcommand{\IZ}{\mathbb{Z}}

\newcommand{\IN}{\mathbb{N}}

\newcommand{\bq}{\begin{equation}}
\newcommand{\eq}{\end{equation}}
\newcommand{\bqa}{\begin{eqnarray}}
\newcommand{\eqa}{\end{eqnarray}}
\newcommand{\ben}{\begin{enumerate}}
\newcommand{\een}{\end{enumerate}}

\newcommand{\ul}{\underline}
\begin{document}

\title{Low SNR Capacity of Fading Channels -MIMO and Delay Spread}
\author{
\authorblockN{Vignesh Sethuraman}
\authorblockA{Qualcomm, Inc.\\
Campbell, California\\
Email: Vignesh.Sethuraman@gmail.com}
\and
\authorblockN{Ligong Wang}
\authorblockA{Signal and Information Processing Laboratory\\
ETH Zurich  \\
Email: wang@isi.ee.ethz.ch}
\and
\authorblockN{Bruce Hajek}
\authorblockA{Department of Electrical and Computer Engineering and the \\
Coordinated Science Laboratory \\
University of Illinois at Urbana-Champaign \\
Email: b-hajek@uiuc.edu}
\and
\authorblockN{Amos Lapidoth}
\authorblockA{Signal and Information Processing Laboratory\\
ETH Zurich  \\
Email: lapidoth@isi.ethz.ch}
}
\maketitle

\begin{abstract}
Discrete-time Rayleigh fading multiple-input multiple-output (MIMO) channels are considered, with no channel state information at the transmitter and receiver. The fading is assumed to be correlated in time and independent from antenna to antenna. Peak and average transmit power constraints are imposed, either on the sum over antennas, or on each individual antenna. In both cases, an upper bound and an asymptotic lower bound, as the signal-to-noise ratio approaches zero, on the channel capacity are presented. The limit of normalized capacity is identified under the sum power constraints, and, for a subclass of channels, for individual power constraints. These results carry over to a SISO channel with delay spread (i.e. frequency selective fading).   
\end{abstract}
\begin{center} \small\bfseries Index Terms \end{center}
Low SNR, channel capacity, correlated fading, frequency selective fading, MIMO, Gauss Markov fading
\vspace{0.67ex}

\section{Introduction}

Discrete-time Rayleigh fading multiple-input multiple-output (MIMO)
channels are considered  in this paper, with no channel side information at the transmitter and receiver.
The fading is assumed to be correlated in time and independent for distinct (input,output)
antenna pairs.  A hard peak constraint, in addition to an average power constraint, is imposed.
The focus of this paper is the low signal to noise ratio (SNR) behavior of the
channel capacity.
Two cases are considered:  either the peak and average power constraints are imposed on the 
sum over the transmit antennas, or
they are imposed on each transmit antenna. In each case, an upper bound on the capacity of MIMO
channels is derived.  In the sum constraint case, the normalized capacity limit as SNR$\to 0$ is identified and the upper bound is found to be tight.   In the individual  constraint case, asymptotic tightness of the upper bound is established under certain conditions.   Insight about optimal signaling strategies is derived in each of the above cases, and comments on the benefits of having multiple antennas in the  low SNR regime are presented.   

 This work summarized here extends previous work of the authors
 \cite{VigneshBH06,LapidothWang06} for  SISO channels, to MIMO and mutlipath fading channels. 
A full length paper in preparation contains the results reported here,
additional bounds for SISO channels, and the proofs.

The capacity of fading channels in the low SNR regime has recently been of much interest \cite{MedardGallager02,HajekSubramanian02,PrelovVerdu04,Laneman05,Collins05,RaoHassibi04,SrinivasanVaranasi06,DurisiBoelcskeiShamai06}.    
For similar work in the high SNR regime, see \cite{KochLapidoth} and references therein.
The main motivation for this work has been to understand the capacity of communication
over wideband channels.  Work of Kennedy \cite{Kennedy69}, Jacobs \cite{Jacobs63},  
Telatar and Tse \cite{TelatarTse00}, and
Durisi et al. \cite{DurisiBoelcskeiShamai06}  demonstrate
that the capacity of such channels, in the wideband limit, is the same as for a wideband
additive Gaussian noise channel with no fading, but the input signals, such as $M$-ary
FSK, are highly bursty in the frequency domain or time domain.  The work of
Medard and Gallager \cite{MedardGallager02} (also see \cite{SubramanianHajek02}) shows
that if the burstiness of the input signals is limited in both time and frequency, then the
capacity of such wideband channels becomes severely limited.  In particular, the required
energy per bit converges to infinity.

Wireless wideband channels typically include both time and frequency selective
fading.   One approach to modeling such channels is to partition the frequency band
into narrow subbands, so that the fading is flat, but time-varying, within each subband.
If the width of the subbands is approximately the coherent bandwidth of the channel,
then they will experience approximately independent fading.   The flat fading
models used in this paper can be considered to be models for communication over
a subband of a wideband wireless fading channel.  The peak power constraints
that we impose on the signals can then be viewed as burstiness constraints in both the
time and frequency domain for wideband communication, similar to those of
\cite{MedardGallager02,SubramanianHajek02}.    However, in this paper, we
consider hard peak constraints, rather than fourth moment constraints as in
\cite{MedardGallager02,SubramanianHajek02}, and we consider the use of multiple
antennas.

The recent work of Srinivasan and Varanasi \cite{SrinivasanVaranasi06} is closely related
to this paper.   It gives low SNR asymptotics of the capacity of MIMO channels with no side
information for block fading channels, with peak and average power constraints,
with the peak constraints being imposed on individual antennas.
One difference between \cite{SrinivasanVaranasi06} and this paper is that
we assume continuous fading rather than block fading. 
In addition, we provide upper bounds on capacity,  and for SISO channels, lower bounds
on capacity, rather than only asymptotic bounds
as in \cite{SrinivasanVaranasi06}.   We assume, however, that the
fading processes are Rayleigh distributed, whereas the asymptotic bounds do not
require such distributional assumption.
The work of Rao and Hassibi \cite{RaoHassibi04} is also related to this paper.
 It gives low SNR asymptotics of the capacity of MIMO channels with no side
information for block fading channels, but the peak constraints are imposed on
coefficients in a particular signal representation, rather than as hard constraints
on the transmitted signals.

Also in this paper, a single-input single-output (SISO) channel with delay spread (i.e. frequency selective) fading is considered. The fading is assumed to be modeled by a finite number of taps.   The fading processes are assumed to be independent across taps, and allowed, within each tap, to be correlated in time.   Lower and upper bounds on the low SNR normalized capacity limit are presented and shown to coincide under some conditions.   The results on SISO channels with delay spread all follow from the results on MIMO channels with individual power constraints.    The model of this section can be thought of as
pertaining to low rate, low power use of a wideband communication channel with constraints on
burstiness in time. 

\section{Preliminaries} \label{sec.preliminaries}

Consider a single-user discrete-time MIMO channel with no channel state information at the
transmitter and receiver. The channel includes additive noise and multiplicative noise
(Rayleigh flat fading). 
Let $N_T$ be the number of  transmit antennas and $N_R$ be the number of receive antennas. 
Let the input at time $n\in \IN$ be denoted by $\sqrt{\rho} \ul{Z}_{N_T \times 1}(n)$: 
so the input on antenna $k$ at time $n$ is $\sqrt{\rho} Z_{k}(n)$. 
Here, the signal to noise ratio is represented by $\rho > 0$. 
Let the corresponding output be denoted by $\ul{Y}_{N_R\times 1}(n)$. Then, 
\begin{equation}
Y_l(n) = \sqrt{\rho} \sum_{k=0}^{N_T-1} {H}_{k,l}(n) Z_{k}(n) + W_{l}(n)
\label{eq:MIMOchModel}
\end{equation}
where $l\in [0,N_R-1]$ is the index of the receive antennas.  
The channel fading processes are assumed to be stationary and ergodic, jointly
proper complex normal (PCN)\footnote{A random vector $Z$ is proper,
in the sense of \cite{NeeserMassey93}, if $E[ZZ^T]=E[Z]E[Z^T]$.  A random process is proper if its
restriction to any finite set of indices gives a proper random vector.  A mean zero PCN
random process is a random process with jointly Gaussian real and imaginary parts,
such that the distribution of the process is invariant under any common rotation of all of its
constituent random variables.},
and spatially independent; 
i.e., if $(k,l) \neq (k',l')$, the fading processes $H_{k,l}$ and $H_{k',l'}$ are mutually independent. 
Further, for each transmit and receive antenna pair $(k,l)$, the fading process  ${H}_{k,l}$ is allowed to be correlated in time,  with autocorrelation function $R_{k,l}(n)$, defined by 
$R_{k,l}(n)= E[{H}_{k,l}(n){H}_{k,l}^*(0)] $, and spectral density function
$(S_{k,l}(\omega))$.
The additive noise on each antenna is modeled by an independent and identically distributed (iid) 
PCN process  with zero mean and unit variance. 
The channel fading processes, the additive noise processes, and the channel input are
assumed to be mutually independent.
The above model does not involve delay spread, but in Section
\ref{sec.delayspread} a SISO delay spread model is examined.

Two different ways to impose peak and average power constraints for MIMO channels
are investigated in this paper.  The constraints are considered either on sums across the
antennas, or on individual antennas.
The {\em sum peak power constraint}  is
\bq
\|\ul{Z}(n)  \|_2^2   \leq 1 ~~\forall~n 
\label{eq:PeakPower}
\eq
and the {\em sum average power constraint} is
\begin{equation}
E[\|\ul{Z}(n) \|_2^2] \le \frac 1 \beta ~~\forall~n,
\label{eq:AvgPower}
\end{equation}
where $\beta \geq 1.$ 
Let $C_{mimo-s}(\rho, \beta)$ denote the information theoretic capacity of the MIMO channel under the sum power constraints (\ref{eq:PeakPower}) and (\ref{eq:AvgPower}). 

 The {\em individual peak power constraints} are
\begin{equation}
|Z_k(n)|^2 \le 1 ~~~ \forall ~ k\in [0,N_T-1]~~~n\in \IZ,
\label{eq:PeakPowerLinf}
\end{equation}
and the {\em individual average power constraints} are
\begin{equation}
E[|Z_k(n)|^2] \le \frac 1 \beta   ~~~  \forall ~ k\in [0,N_T-1]~~~n\in \IZ,
\label{eq:AvgPowerPerAnt}
\end{equation}
where $\beta \geq 1$.
Such constraints are seen in practice when each transmit antenna is powered by its own analog driver,
so, at any time instant, the available instantaneous power for each antenna is not immediately constrained by the instantaneous powers of the other antennas. 
Let $C_{mimo-i}(\rho,\beta)$ denote the information theoretic capacity of the channel under the
individual power constraints (\ref{eq:PeakPowerLinf}) and (\ref{eq:AvgPowerPerAnt}).

Two constants,  $\phi_{k,l}$ and $\lambda_{k,l}$, and a function, $I_{k,l}$,
are associated with the autocorrelation function
$R_{k,l}$ of a fading process $H_{k,l}=(H_{k,l}(n): n \in \IZ)$, as follows.
The constants are defined by
\bq
\phi_{k,l} = \sum_{n=1}^{\infty} |R_{k,l}(n)|^2~~~\mbox{and}~~~
\lambda_{k,l}  = (R_{k,l}(0))^2 + 2\phi_{k,l}  .
\eq
It is assumed that $\phi_{k,l}$ is finite for all valid $k$ and $l$.  
For SISO channels, $R(n)$, $\phi$, and $\lambda$, are similarly determined
by the single fading process $H.$
The function is defined by
$$
I_{k,l}(\rho) = \int_{-\pi}^{\pi} \log(1+\rho S_{k,l}(\omega)) \frac {d\omega}{2\pi}.
$$
An interpretation of $I_{k,l}(\rho)$ is that it is the mutual information rate between the random process
$(H_{k,l}(n): n \in \IZ)$ and a random process of the form $(\sqrt{\rho} H_{k,l}(n) + W(n): n \in \IZ)$,
where $W$ is an iid PCN sequence with unit variance, independent of $H_{k,l}$.
%
%
A MIMO channel is said to be
 {\em transmit separable} if there are nonnegative constants $(\alpha_k: 0\leq k \leq N_T-1)$
and autocorrelation functions $(R_{l} : 0 \leq l \leq N_R-1)$ so that
$R_{k,l}(n) = \alpha_k R_l(n)$ for all $k,l,$ and $n$.  In this case we have
$\phi_{k,l} = \alpha_k^2 \phi_{l},$  $\lambda_{k,l}=\alpha_k^2\lambda_l$, and
 $I_{k,l}(\rho)=I_l(\rho\alpha_k)$,
where $\phi_l$, $\lambda_l$, and $I_l$ are associated with the
autocorrelation function $R_l$ for each $l$.

We call an individual fading process $H_{k,l}$  {\em ephemeral} if
$  2\phi_{k,l} \le   R_{k,l}^2(0)$ and {\em nonephemeral} otherwise.
A MIMO channel is said to be {\em nonephemeral} if all of the constituent fading
processes $\left\{ H_{k,l} \right\}$ are nonephemeral.

\section{MIMO Channels with Sum  Constraints}
\label{sec.SumPeakConstraintMainResults}

This section concerns the MIMO channel with the power constraints
\eqref{eq:PeakPower} and  \eqref{eq:AvgPower}
on sums over antennas.     Let
\begin{equation}
{\cal A}(\beta) = \left\{(a_0, \dots, a_{N_T-1}): a_k \geq 0 ~\forall ~k, ~\sum_{k=0}^{N_T-1} a_k \leq \frac 1 \beta \right\},
\label{eq:setA}
\end{equation}
and
\begin{eqnarray*}
U_{mimo-s}(\rho, \beta) = \max_{\ul{a} \in {\cal A(\beta)}}
\sum_{l=0}^{N_R-1} 
\left\{
\log(1+ \rho \sum_{k=0}^{N_T-1} R_{k,l}(0) a_k)    \right.  \nonumber
\\ \left.
- \sum_{k=0}^{N_T-1} a_k I_{k,l}(\rho) \right\}.~~~~~~
\label{eq:mimoUB}
\end{eqnarray*}
\begin{prop}  \label{prop:mimo_s_U}
$C_{mimo-s}(\rho, \beta) \le U_{mimo-s}(\rho, \beta)$.
\end{prop}

The following result identifies the asymptotic behavior of $C_{mimo-s}(\rho, \beta)$ at
low $\rho$ (for a fixed $\beta$).

\begin{prop} \label{prop:mimo_s_asymp}
For $\beta \geq 1 $ fixed, 
\begin{eqnarray*}
\lefteqn{ \lim_{\rho \to 0}
\frac{C_{mimo-s}(\rho, \beta)}{\rho^2} =  \lim_{\rho \to 0}
\frac{U_{mimo-s}(\rho, \beta)}{\rho^2} } \\
&=& \frac 1 2 \max_{\ul{a} \in {\cal A(\beta)}}
\sum_{l=0}^{N_R-1}
\left\{ \sum_{k=0}^{N_T-1} a_k\lambda_{k,l}  \right.\\
&&  ~~~~~~~~~~~~~~~  -  \left.   \left(\sum_{k=0}^{N_T-1} a_k R_{k,l}(0)\right)^2 \right\}.
\end{eqnarray*}
\end{prop}

The input distribution used in the proof of the asymptotic lower bound portion
of Proposition \ref{prop:mimo_s_asymp}  has the following form.
For $\ul{a} \in {\cal A}(\beta)$, the input $(\underline{Z}(1), \ldots , \underline{Z}(n))$
can be represented as follows.  For $1\leq n \leq N$,
\begin{equation}
Z_0(n) = \mathbf{1}_{\{U \le a_0\}} \exp(j \theta(n))
\nonumber 
\end{equation}
and for $0 < k   < N_T$,
\begin{equation}
Z_k(n) = \mathbf{1}_{\{\sum_{i=0}^{k-1}a_i \le U \le \sum_{i=0}^{k} a_i\}} \exp(j \theta(n)),
\nonumber 
\end{equation}
where $j=\sqrt{-1},$ $\mathbf{1}_A$ represents the indicator function of an event $A$,
$U$ is uniformly distributed on the interval $[0,1]$, and
the phases $\theta(1), \ldots , \theta(n)$
can be chosen in any one of the following ways:
\begin{enumerate}
\item $\theta(n) = n \vartheta$,
where $\vartheta$ is a discrete random variable, uniformly distributed over
$\{\frac {2\pi l}N: 0\le l\leq N-1\}$. 
This is a form of frequency shift keying (FSK), related to the M-FSK modulation presented in
\cite{Viterbi67} in a continuous time setting.
\item
Or, $\theta(n) = n \vartheta$,
where $\vartheta$ is uniformly distributed over $[0, 2\pi].$ This is a limiting form of FSK for the
number of tones going to infinity.
\item Or,  $\theta(n)$, $1\leq n \leq N$, are independent, with $\theta(n)$ for each $n$ being
uniformly distributed over $\{ 2\pi i/d: 0 \leq i \leq d-1\}$ (i.e. $d$-ary phase shift keying) for some integer $d\geq 2$, or uniformly distributed over $[0,2\pi]$.
\end{enumerate}
Thus for this input distribution, at most one antenna is used at any time instant. 
Moreover, the same antenna (if any) is used for all of the $N$ channel uses.
Antenna $k$ is used with probability $a_k$.

\begin{cor} \label{mimo_s_Xsep}
 If the MIMO channel is transmit separable, then $C_{mimo-s}(\rho,\beta)$
 is bounded from above by
\bq     \label{eq:mimo_s_XsepU}
\max_{0\le a \le \frac 1 \beta}  \sum_{l=0}^{N_R-1} \left\{
\log(1 + a \alpha_{\max} \rho R_l(0) ) - aI_l(\alpha_{\max}\rho) \right\}
\eq
where $\alpha_{\max}=\max\{\alpha_0, \ldots , \alpha_{N_T-1} \}.$
The bound \eqref{eq:mimo_s_XsepU} is asymptotically tight as $\rho \to 0$, and
\bq         \label{eq:mimo_s_XsepAsymp}
\lim_{\rho \to 0} \frac{C_{mimo-s}(\rho, \beta)}{\rho^2} =
\frac {\alpha_{\max}^2}{2}  \max_{0\leq a \leq \frac{1}{\beta}  }
\sum_{l=0}^{N_R-1}
\left\{ a \lambda_l
- a^2 R_l^2(0) \right\}.
\eq
 \end{cor}
The reasoning behind Corollary \ref{mimo_s_Xsep} is that,
for a transmit separable channel with sum constraints, the asymptotic capacity can be
achieved by using only one transmit antenna $k$ with the largest $\alpha_k$, and sending
the same signal on it as for SISO channels.
The next corollary is a simple special case of Corollary \ref{mimo_s_Xsep}:
\begin{cor}
\label{cor.mimo_s_Xsep_noneph}
If the channel is transmit separable and nonephemeral,  and if no average
power constraint is imposed (i.e.  $\beta=1$), then
$$
\lim_{\rho \to 0} \frac{C_{mimo-i}(\rho,1)}{\rho^2}
=  \alpha_{\max}^2   \sum_{l=0}^{N_R-1} \phi_l.
$$
\end{cor}
As the proof indicates, under the conditions of Corollary \ref{cor.mimo_s_Xsep_noneph},
the optimal input is to use only one antenna $k$ with maximum $\alpha_k$,
and send on it a signal of the form $Z_k(n)=\exp(j\theta(n))$, with the sequence
$\theta(n)$ selected as before.  This input distribution is the same found to be optimal
by  \cite{RaoHassibi04} for block fading channels and a different type of peak constraint,
although in \cite{RaoHassibi04} the channels are assumed to be statistically identical, so
that any of the transmit antennas could be used.

\section{MIMO Channels with Individual Constraints}
\label{sec.IndivConstraints}

This section concerns MIMO channels with the individual power constraints
\eqref{eq:PeakPowerLinf} and  \eqref{eq:AvgPowerPerAnt}.
In the full version of this paper, an upper bound and an asymptotic lower bound
on the capacity are given.  The bounds are not included here for lack of space. In general, 
the asymptotic bounds are not equal.  However, for the broad class of transmit
separable channels, the normalized capacity limit can be identified:

\begin{cor} \label{cor.mimoXsep}
If the channel is transmit separable, then
$\lim_{\rho \to 0} \frac{C_{mimo-i}(\rho,\beta)}{\rho^2} $ is equal to
$$
 \frac 1 2 \left( \sum_{k=0}^{N_{T-1}} \alpha_k\right)^2
 \max_{0 \le a \le \frac 1 \beta} \sum_{l=0}^{N_R-1} \left\{ a\lambda_l
- a^2  R_l^2 (0) \right\}.
$$
\end{cor}

The input strategy used to obtain the asymptotic lower
bound portion of Corollary \ref{cor.mimoXsep} is to transmit the same signal on all antennas,
with the signal on each antenna having the distribution described above for a single antenna
in the sum constraint case.   This is exactly the distribution, called STORM, proposed for
block fading channels by Srinivasan and Varanasi \cite{SrinivasanVaranasi06}.

The next corollary is a simple special case of Corollary \ref{cor.mimoXsep}:
\begin{cor}
\label{cor.mimo_i_Xsep_noneph}
If the channel is transmit separable and nonephemeral,  and if no average
power constraint is imposed (i.e.  $\beta=1$), then
$$
\lim_{\rho \to 0} \frac{C_{mimo-i}(\rho)}{\rho^2}
=  \left( \sum_{k=0}^{N_{T-1}} \alpha_k\right)^2  \sum_{l=0}^{N_R-1} \phi_l.
$$
\end{cor}

The next corollary looks at the general bounds in another direction, although the resulting
upper and lower bounds do not match.
\begin{cor}
\label{cor:mimoEphemAsympLInf}
If the channel is  nonephemeral and if no average power constraint is imposed
(i.e.  $\beta=1$), 
\begin{equation}
\lim\sup_{\rho \to 0} \frac{C_{mimo-i}(\rho)}{\rho^2} \leq
 N_T \sum_{k,l} \phi_{k,l} 
\label{eq:mimoAsympEphemUBLInf}
\end{equation}
and 
\begin{equation}
\lim\inf_{\rho \to 0} \frac{C_{mimo-i}(\rho)}{\rho^2} \geq
\sum_{l=0}^{N_R-1}  \sum_{n=1}^{\infty}
\left|\sum_{k=0}^{N_T-1}  R_{k,l}(n) \right|^{2}.
\label{eq:mimoAsympEphemLBLInf}
\end{equation}
\end{cor}
The input strategy used to obtain the asymptotic lower
bound portion of Corollary \ref{cor:mimoEphemAsympLInf}  is to use $Z_k(n)=\exp(j\theta(n))$, with
the phase sequence chosen as for the SISO channel.  So all antennas send the same constant
magnitude signal.

\section{Discussion of results for MIMO channels}

Some remarks on the above results for MIMO channels are given next.
We first comment on the benefits of having multiple antennas,
the input distributions that achieve the lower asymptotic bounds, and
the relationship between channel memory and capacity.
For simplicity, consider the low SNR normalized capacity limit for a nonephemeral
MIMO channel with peak constraints but no average power constraints.
For individual peak constraints, the limit is, according to 
Corollary \ref{cor.mimo_s_Xsep_noneph},
$\alpha_{\max}^2 \sum_{l=0}^{N_R-1} \phi_l$, which does not
grow with the number of transmit antennas, as long as the $\alpha$ 
values for additional antennas are not larger than the $\alpha$ of the
first antenna.  The intuitive reason is that any benefit due to diversity brought by
multiple transmit antennas is nulled by the cost of tracking the
additional fading processes.

If, for the same channel, individual peak constraints are imposed, Corollary
\ref{cor.mimo_i_Xsep_noneph} yields that the normalized capacity limit is
$\left(  \sum_{k=0}^{N_T-1} \alpha_k\right)^2 \sum_{l=0}^{N_R-1} \phi_l.$
This is larger than for the sum constraint case because of the
following two facts:  (i) the average received
power is a factor  $\left(  \sum_{k=0}^{N_T-1} \alpha_k\right)/\alpha_{\max}$
larger for the individual constraint case and
 (ii) the normalized capacity limit, obtained by dividing by
$\rho^2$, scales quadratically with an effective factor change in the SNR, $\rho$.

The peak power constraints can be adjusted in the two cases to yield the same
maximum  total transmitted power by replacing $\rho$ by $N_T\cdot  \rho$ in the case
of sum power constraints.  Then, the normalized capacity limit for the sum
power constraints is $(N_T\alpha_{\max})^2 \sum_{l=0}^{N_R-1} \phi_l$, which is
larger than the limit for the case of individual constraints, unless the $\alpha_k$'s are all
equal.  If the $\alpha_k$'s are equal, the normalized capacity limit is the same
whether the peak constraint is
applied as a sum peak constraint or as individual peak constraints per antenna.
In certain applications,  it may be cheaper to produce multiple transmit antennas,
each with a small peak power capability, than a single transmit antenna that
provides a proportionately larger capability.
In such a case, this analysis tilts the balance in favor of using multiple transmit antennas.

Having multiple receive antennas is tremendously useful at low SNR, for either peak or sum
constraints.  The amount of information learned about the input by each receive antenna is so
small (at low SNR) that each additional receive antenna gathers information that is almost independent of that gathered by the other antennas.  Therefore, as can be seen in the normalized capacity
limits, the channel capacity is linear in the number of receive antennas at low SNR.

\section{SISO Channels with Delay Spread}
\label{sec.delayspread}

Consider the following SISO fading channel model with delay spread:
\bq
Y(n) = \sqrt{\rho} \sum_{k=0}^{K-1} H_k(n) Z(n-k) + W(n)
\label{eq:multipathModel}
\eq
Here, $Z$ is the input (complex scalar) and $Y$ is the output.
The fading is assumed to be independent across the $K$ taps, and correlated in time within each tap. The correlation function for tap $k$ is given by
$\{R_k(n): n\in \IZ\}$.
The additive noise is modeled by $W$, an iid PCN process with zero mean and unit variance, and is assumed to be independent of the fading and the input.

We assume the input is subject to the same constraints as considered earlier for SISO channels
with flat fading:  the peak constraint is $|Z(n)| \leq 1$ for all $n$, and the average power constraint
is  $E[|Z(n)|^2] \leq 1/\beta$ for all $n$, for some constant $\beta \geq 1$.
We denote the capacity of the delay spread channel by $C_{ds}(\rho, \beta)$.

The channel in (\ref{eq:multipathModel}) is equivalent to a MISO channel with 
$N_T=K$ antennas, and an additional constraint.  The MISO channel is given by:
$$
Y(n) = \sqrt{\rho} \sum_{k=0}^{K-1} H_k(n) Z_k(n) + W(n),
$$
with the process $H_k$ representing the fading process for antenna $k$, 
and the vector input $\underline{Z}(n)$ for the MISO channel being given
by  $Z_k(n)  = Z(n-k)$  for $0 \leq k \leq N_T-1.$   The peak and average power
constraints on the SISO delay spread channel imply that peak and average power
constraints are satisfied for the individual antennas of the MISO system. 
The additional constraint required for a MISO input to correspond to a SISO
input is:
\bq  \label{eq:MISO-DSconstraint}
Z_k(n)=Z_{k'}(n') ~ \mbox{whenever} ~ n-k=n'-k'.
\eq
Consequently, $C_{ds}(\rho,\beta) \leq C_{miso-i}(\rho,\beta)$, with the understanding that
the fading processes for the antennas of the MISO channel are the fading processes of the
taps of the SISO delay spread channel.  This inequality is not, in general, tight, because
$C_{miso-i}(\rho,\beta)$ is the capacity of the MISO channel without the extra
constraint \eqref{eq:MISO-DSconstraint}.

The bounds and asymptotics for $C_{mimo-i}$, given is Section \ref{sec.IndivConstraints},
can be easily specialized to MISO channels by taking $N_R=1$, and dropping the subscript
$l$ indexing receive antennas.  The upper bounds carry over to the SISO delay spread
channel without change.  (It is possible tighter bounds could be obtained by incorporating
the constraint \eqref{eq:MISO-DSconstraint}, but we don't pursue that here.)  The lower
bounds carry over to the extent that the inputs they are based on satisfy the extra constraint
\eqref{eq:MISO-DSconstraint}.   Upper bounds and asymptotic lower bounds
are presented in the full version of this paper.  Here we present corollaries
for two subclasses of channels.

A SISO delay spread channel is said to be {\em delay separable} if
there are nonnegative constants $\alpha_0, \ldots, \alpha_{K-1}$
and an autocorrelation function $R$ so that
$R_k(n)=\alpha_k R(n)$ for $0 \leq k \leq K-1$.
\begin{cor} \label{cor.dssep}
If the channel is delay separable, then
\begin{eqnarray*}
\lim_{\rho \to 0} \frac{C_{ds}(\rho,\beta)}{\rho^2} =
 \frac 1 2 \left( \sum_{k=0}^{K-1} \alpha_k\right)^2
 \max_{0 \le a \le \frac 1 \beta} \left\{ a\lambda
- a^2  R^2 (0) \right\}.
\end{eqnarray*}
\end{cor}
The next corollary is a simple special case of Corollary \ref{cor.dssep}:
\begin{cor}
\label{cor.ds_delaysep_noneph}
If the channel is delay separable and nonephemeral,  and if no average
power constraint is imposed (i.e.  $\beta=1$), then
$$
\lim_{\rho \to 0} \frac{C_{ds}(\rho)}{\rho^2}
=  \left( \sum_{k=0}^{K-1} \alpha_k\right)^2\phi.
$$
\end{cor}
\begin{cor}
\label{cor:dsEphemAsympLInf}
If the channel is  nonephemeral and if no average power constraint is imposed
(i.e.  $\beta=1$) then
$$
\lim\sup_{\rho \to 0} \frac{C_{ds}(\rho)}{\rho^2} \leq K \sum_{k=0}^{K-1} \phi_{k} 
$$
and 
$$
\lim\inf_{\rho \to 0} \frac{C_{ds}(\rho)}{\rho^2} \geq
 \sum_{n=1}^{\infty}
\left|\sum_{k=0}^{K-1}  R_{k}(n) \right|^{2}.
$$
\end{cor}

As mentioned earlier in the subsection, the input distribution used to prove the asymptotic
lower bound portions of the bounds above is an $N$-ary FSK signal with the addition of the
all zero signal.  Although this is the same distribution that can be used to achieve the
same capacity as for an AWGN channel
\cite{Kennedy69,Jacobs63,TelatarTse00,DurisiBoelcskeiShamai06}, the fact that
we consider the normalized limit as the peak transmit power for the whole wideband channel
converges to zero,
results in a lower spectral efficiency, with capacity tending to zero quadratically in
$\rho$.

\section*{Acknowledgment}
This work was supported in part by the National Science Foundation Grant NSF ITR 00-85929,
and the Motorola Center for Communication Graduate Fellowship at UIUC.  
This work was conducted while V. Sethuraman was at UIUC.

\bibliographystyle{ieeetr}

\end{document}